\documentclass[nofootinbib,prx,twocolumn,showpacs,superscriptaddress,notitlepage,superscriptaddress,amsmath]{revtex4-2}

\usepackage{lipsum}  
\usepackage{dsfont}  
\usepackage{amsmath}
\usepackage{soul}
\usepackage{braket}
\usepackage{amssymb}
\usepackage{amsthm}
\usepackage{algpseudocode}
\usepackage{algorithm}
\usepackage{siunitx}
\usepackage{amsfonts}
\usepackage{comment}
\usepackage[normalem]{ulem}
\usepackage{graphicx}
\usepackage{color,framed}
\usepackage{hyperref}
\usepackage{enumerate}
\usepackage{lipsum}
\usepackage{slashed}
\usepackage{url}
\usepackage{bbm}
\usepackage{chngcntr}
\counterwithout{equation}{section}
\usepackage{tikz,pgfplots}
\usepackage{pgfplotstable}
\usepackage{siunitx}
\usepackage{comment}
\usepackage{graphicx}

\makeatletter
\newcommand{\setlabel}[1]{\edef\@currentlabel{#1}\label}
\makeatother

\usepgfplotslibrary{fillbetween}

\definecolor{keigreen}{rgb}{0,0.5,0}

\hypersetup{
    colorlinks=true, 
    linktoc=all,     
    linkcolor=blue,  
}

\def \beq {\begin{equation}}
\def \eeq {\end{equation}}
\def \beqa {\begin{eqnarray}}
\def \eeqa {\end{eqnarray}}
\def \bseq {\begin{subequations}}
\def \eseq {\end{subequations}}

\newcommand \tr {{\rm tr}\,}

\newcommand{\D}[1]{\text{d}#1}

\pgfplotsset{compat=1.18}
\begin{document}

\title{Dynamical structure factor with a pumping approach \\
on a trapped-ion quantum computer}

\author{Etienne Granet}
\affiliation{Quantinuum, Leopoldstrasse 180, 80804 Munich, Germany}

\author{Keisuke Murota}
\affiliation{Department of Physics, The University of Tokyo, Tokyo 113-0033, Japan}

\author{Henrik Dreyer}
\affiliation{Quantinuum, Leopoldstrasse 180, 80804 Munich, Germany}

\author{Kentaro Yamamoto}
\affiliation{Quantinuum K.K., Otemachi Financial City Grand Cube 3F, 1-9-2 Otemachi, Chiyoda-ku, Tokyo, Japan}

\author{Juan Pedersen}
\affiliation{Quantinuum K.K., Otemachi Financial City Grand Cube 3F, 1-9-2 Otemachi, Chiyoda-ku, Tokyo, Japan}

\author{Hidemaro Suwa}
\affiliation{Department of Physics, The University of Tokyo, Tokyo 113-0033, Japan}

\date{\today}
\begin{abstract}
Dynamical structure factors (DSF) measured with neutron-scattering experiments provide key insights into the structure of materials. Their computation requires both the preparation of an equilibrium state and the implementation of Hamiltonian dynamics. We demonstrate the feasibility of computing DSF on the Quantinuum Reimei trapped-ion quantum computer, comparing the DSF of 1D Heisenberg model on $20$ sites, and that of the copper sulfate crystal. To that end, we introduce a pumping approach for computing the DSF $S(q,\omega)$ on quantum computers that enables targeting specific arbitrary values of frequencies $\omega$. This method time-evolves the initial state using a time-dependent Hamiltonian perturbed by a source term oscillating at the target frequency \(\omega \). When targeting only a few frequency values, this approach provides a significant reduction in shot overhead compared to previous methods.

\end{abstract}
\maketitle

\textbf{\emph{Introduction.}}--- Neutron scattering experiments are a common technique used in condensed matter physics to probe the nature of materials. These experiments involve irradiating a sample with a neutron beam and measuring the post-scattering momentum and energy using an array of detectors~\cite{sturm1993dynamic}. The number of neutrons measured with an energy difference $\omega$ and a momentum difference $q$ compared to the initial beam is then directly proportional to $S(q,\omega)$, the dynamical structure factor (DSF) of the material. This quantity can be written as the Fourier transform, both in space and time, of the dynamical correlations $\langle \mathcal{O}(x,t)\mathcal{O}(0,0) \rangle$ of some observable $\mathcal{O}(x,t)$ localized at point $x$ and evolved for time $t$ with the Hamiltonian $H$ describing the system \cite{Ashcroft76}. The expectation value is typically taken in a thermal state at some inverse temperature $\beta$ or in the ground state of $H$.

Although the regime of weak quantum fluctuations can be classically addressed with spin-wave theory \cite{fishman2018spin} and semi-classical approaches \cite{Schulz90,chern18,mazzone22,dahlbom22,yamasaki26}, the computation of these dynamical correlations is in general a difficult problem for classical computers. This difficulty is twofold, (i) one must find some representation of the ground state or some thermal state, and (ii) one must compute Hamiltonian dynamics on top of a perturbation, which are both difficult tasks for classical computers in general. In contrast, quantum computers can prepare finite-temperature states and perform Hamiltonian dynamics efficiently \cite{rajakumar2026gibbs,bergamaschi2024quantum,lloyd1996universal}. Approximating dynamical structure factors is also known to be BQP-hard \cite{baez2020dynamical}. 

Given a state preparation technique, several approaches to compute $S(q,\omega)$ on a quantum computer have been developed. Many methods compute dynamical correlations with Hamiltonian dynamics and use variants of a Hadamard test \cite{somma2002simulating,knap2013probing,chiesa2019quantum,roggero2019dynamic,francis2020quantum,kosugi2020construction,tacchino2020quantum,baez2020dynamical,endo2020calculation,libbi2022effective,altuntas2024dynamical,kokcu2024linear,guimaraes2025accelerating,yi2025ancilla,cruz2025quantum,piccinelli2026circuit}. These approaches all require to compute the dynamical correlations individually for every time point, and then to compute their Fourier transform. There are multiple hardware implementations of these techniques on various platforms \cite{chiesa2019quantum,kastner2024ancilla,eassa2024high,bauer2025progress,vilchez2025extracting,sun2025probing,lee2026benchmarking}.  There also exist variational methods \cite{chen2021variational,huang2022variational,gyawali2022adaptive,lee2022variational,jensen2023near}, and methods working directly in momentum space \cite{ciavarella2020algorithm,keen2021quantum}.

In this work, we introduce a pumping method where for thermal states, the DSF $S(q,\omega)$ is directly obtained by measuring $\mathcal{O}(x)$ on the quantum computer, without the intermediary of dynamical correlations. The method avoids computing the dynamical correlations for several different time points, and is able instead to directly target specific  $\omega$ frequencies. This is particularly efficient if only a few frequencies are required. We implement the algorithm on the Quantinuum Reimei trapped-ion quantum computer within the Fugaku--Reimei hybrid quantum--HPC environment, comparing the DSF of 1D Heisenberg model on $20$ sites with noiseless simulations, and the DSF of the copper sulfate crystal.

\textbf{\emph{Pumping approach.}}--- Given a Hamiltonian $H$ and an observable $\mathcal{O}$, we define the DSF as
\begin{equation}\label{fourier}
    S(q,\omega)=\int_{-\infty}^\infty \D{t} \int\D{x} \, \langle \mathcal{O}(x,t)\mathcal{O}(0,0) \rangle e^{i(\omega t-qx)}\,,
\end{equation}
where $\langle \cdot \rangle=\tr[\rho \cdot]$ denotes the expectation value in the density matrix $\rho=\frac{e^{-\beta H}}{\tr[e^{-\beta H}]}$, and $\mathcal{O}(x,t)=e^{iHt}\mathcal{O}(x)e^{-iHt}$. Our setting applies to any dimension, but on a 1D lattice the integration $\int\D{x}$ must be understood as a sum over sites $\frac{1}{L}\sum_{x=0}^{L-1}$, where $L$ is the number of sites.

Let us fix a final time $T$ and a frequency $\omega$. We define the following time-dependent Hamiltonian
\begin{equation}
    H(t)=H-\epsilon \mathcal{O}(0) \sin(\omega (T-t))\,,
\end{equation}
for some coupling $\epsilon$, and measure $\mathcal{O}(x)$ after evolving for time $t$, namely $\langle \mathcal{O}(x,t)\rangle_\epsilon\equiv\tr[\rho W^\dagger(t) \mathcal{O}(x) W(t)]$ where
\begin{equation}
    W(t)=\mathcal{T}\exp \left(-i\int_0^t \D{s}H(s)\right)\,.
\end{equation}
Expanding in $\epsilon$, we have
\begin{equation}
\begin{aligned}
    W(T)=&e^{-iTH}+i\epsilon \int_0^T \D{s}\sin(\omega(T-s)) e^{-iH(T-s)}\mathcal{O}(0)e^{-iHs}\\
    &+\mathcal{O}(\epsilon^2)\,.
\end{aligned}
\end{equation}
We get
\begin{equation}
\begin{aligned}
    &\langle \mathcal{O}(x,T)\rangle_\epsilon-\langle \mathcal{O}(x,T)\rangle_{-\epsilon}\\
    &=4i\epsilon \int_{0}^T \D{s} \langle \mathcal{O}(x,s)\mathcal{O}(0,0)\rangle \sin (\omega s)+\mathcal{O}(\epsilon^3)\\
    &=2i\epsilon \int_{-T}^T \D{s} \langle \mathcal{O}(x,s)\mathcal{O}(0,0)\rangle \sin (\omega s)+\mathcal{O}(\epsilon^3)\,,
\end{aligned}
\end{equation}
where we used that the averaging state is an equilibrium state, and where we assumed translation invariance and $x\to -x$ symmetry of the Hamiltonian to obtain the last line. Then, defining
\begin{equation}
    \hat{\mathcal{O}}_{q}=\int\D{x}\, \mathcal{O}(x) \cos(q x)\,,
\end{equation}
we have
\begin{equation}
\begin{aligned}
    &S(q,\omega)-S(q,-\omega)\\
    &=\underset{T\to\infty}{\lim}\frac{\langle \hat{\mathcal{O}}_{q}(T)\rangle_\epsilon-\langle \hat{\mathcal{O}}_{q}(T)\rangle_{-\epsilon}}{\epsilon}+\mathcal{O}(\epsilon^2)\,.
\end{aligned}
\end{equation}
Since for a thermal state, we have the detailed balance relation $S(q,-\omega)=e^{-\beta\omega}S(q,\omega)$, we conclude that we can write
\begin{equation}\label{dsf}
    S(q,\omega)=\underset{T\to\infty}{\lim}\frac{\langle \hat{\mathcal{O}}_{q}(T)\rangle_\epsilon-\langle \hat{\mathcal{O}}_{q}(T)\rangle_{-\epsilon}}{\epsilon (1-e^{-\beta\omega})}+\mathcal{O}(\epsilon^2)\,.
\end{equation}

We note that with this method, to compute the DSF we only need to measure local operators $\mathcal{O}(x)$ after evolving the system with a time-dependent Hamiltonian.

\textbf{\emph{Concrete application to a 1D model.}}--- 
We are going to apply this pumping approach to compute the DSF of the $1D$ Heisenberg model with periodic boundaries
\begin{equation}
    H=\sum_{j=0}^{L-1}X_jX_{j+1}+Y_jY_{j+1}+Z_jZ_{j+1}\,.
\end{equation}
This model is interesting for multiple reasons. Firstly, it is known to describe crystals of copper sulfate CuSO${}_4$ very well: neutron scattering experiments have been performed to compute the dynamic structure factor of that material, with excellent agreement with that of the 1D Heisenberg model \cite{mourigal2013fractional}. There are thus already neutron scattering experiments at our disposal to compare with the results of the quantum hardware. Secondly, the model is exactly solvable with the Bethe ansatz~\cite{bethe1931theorie}, and advanced techniques allow for computing the dynamic structure factor for large number of sites, which is relevant for benchmarking purposes \cite{caux2009correlation}. Lastly, the 1D nature of the system also allows one to simulate larger sample sizes and finer momentum resolution. We note that this same model has been recently used for demonstrating the computation of DSF on quantum computers by other groups \cite{lee2026benchmarking}.

Specifying the general framework of the previous section to this 1D model, we define thus
\begin{equation}\label{hatx}
    \hat{X}_q=\frac{1}{L}\sum_{j=0}^{L-1}X_j \cos(qj)\,,
\end{equation}
for $q=\frac{2\pi n}{L}$ a quantized one-dimensional momentum with $n=0,...,L-1$. Formula \eqref{dsf} is then
\begin{equation}\label{dsf1d}
\begin{aligned}
    S(q,\omega)&=\underset{T\to\infty}{\lim}\frac{\langle \hat{X}_q(T)\rangle_\epsilon-\langle \hat{X}_q(T)\rangle_{-\epsilon}}{\epsilon(1-e^{-\beta\omega})}+\mathcal{O}(\epsilon^2)\,.
\end{aligned}
\end{equation}
 Our time-dependent Hamiltonian is
\begin{equation}\label{hamiltonian}
    H(t)=H-\epsilon X_0 \sin(\omega (T-t))\,.
\end{equation}

\begin{figure}
    \centering
    \includegraphics[width=0.9\linewidth]{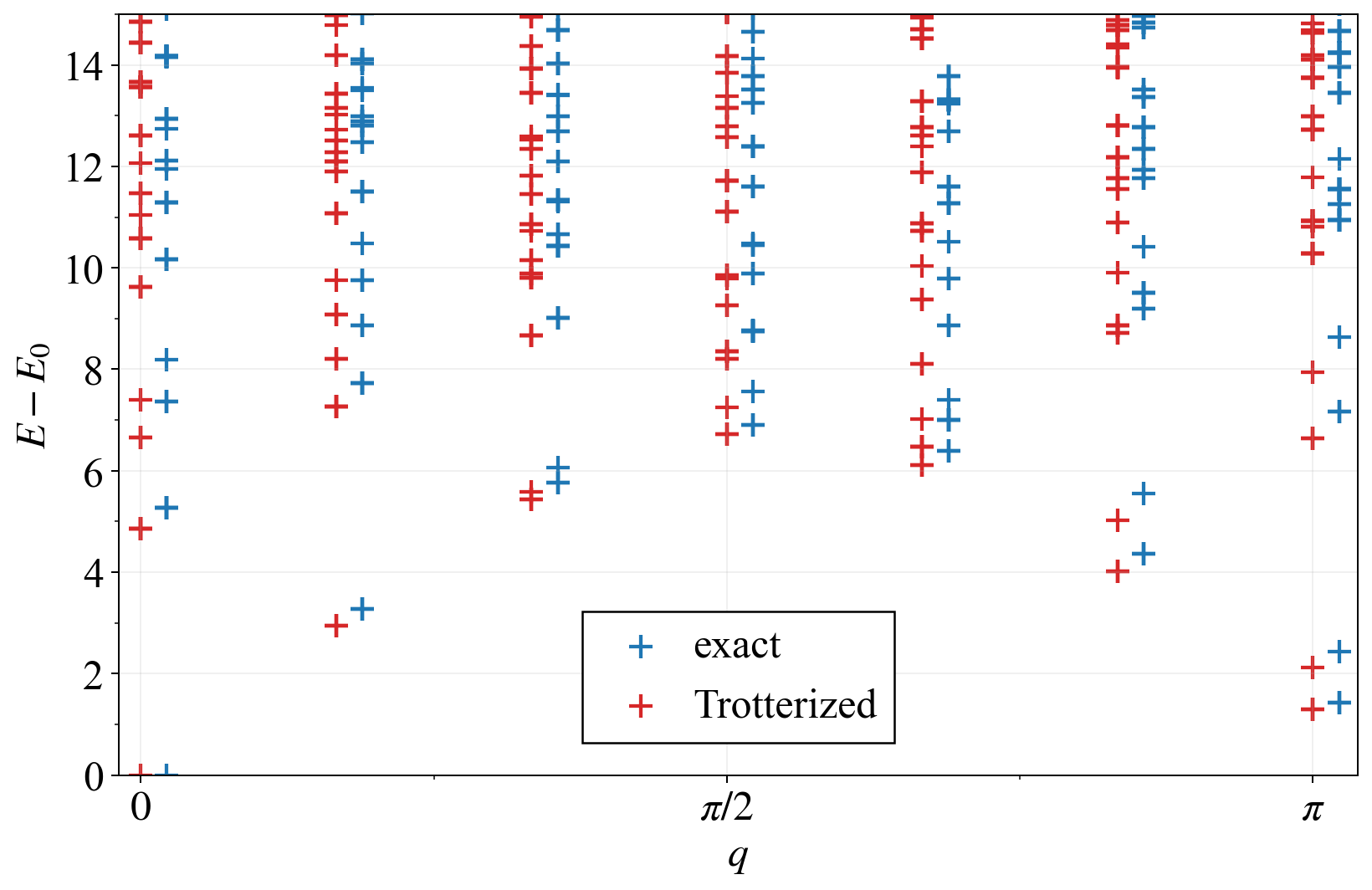}
    \caption{\emph{Influence of the approximate time evolution.} The position in the energy-momentum space of the excitation energies of the 1D Heisenberg model in size $L=12$, from the exact Hamiltonian $H$ (blue), and from the corresponding Floquet Hamiltonian $H_F$ in \eqref{floquet} with ${\rm d}t=0.24$ (red).
    The momenta of the exact Hamiltonian are slightly shifted for better visibility.
    }
    \label{fig:time}
\end{figure}

Formula \eqref{dsf1d} requires to take the limit $T\to\infty$. In an actual implementation, this limit cannot obviously be taken exactly. A natural regularization scheme on a quantum computer is to truncate the time evolution to some finite large enough $T$. This corresponds to truncating the time integral in \eqref{fourier} to values between $-T$ and $T$. The dynamical correlation $\langle X_0X_j(t)\rangle$ can be written as sums of eigenstates $|E_n\rangle$ of the Hamiltonian $H$
\begin{equation}
    \langle X_j(t)X_0\rangle=\sum_{n} |\langle E_0 | X_0 | E_n \rangle|^2 e^{it(E_0-E_n)}e^{ij(P_n-P_0)}\,,
\end{equation}
with $E_n$ the energy and $P_n$ the momentum, with index $0$ denoting the ground state. Truncating the time integral to a finite $T$ gives the approximate DSF $\tilde{S}(q,\omega)$
\begin{equation}\label{lehman}
    \tilde{S}(q,\omega)=\sum_{P_n-P_0=q} |\langle E_0 | X_0 | E_n \rangle|^2 \frac{\sin [ T(E_n-E_0-\omega)]}{ T (E_n-E_0-\omega)}\,.
\end{equation}
Taking $T\to\infty$ imposes that $E_n-E_0=\omega$. In a finite system, there is always a finite number of states, so that $\tilde{S}(q,\omega)$ becomes a finite sum of $\delta$ functions. Taking a finite $T$, besides making the computations on the quantum computer shorter, smoothens the DSF obtained. The sinc function can however cause spurious bumps to appear.

Another source of approximation is that the time evolution needs to be expressed in terms of gates with some algorithm. We will use a Trotter decomposition~\cite{suzuki1990fractal} and write
\begin{equation}\label{trotter}
\begin{aligned}
W(T)\approx&\prod_{m=1}^{T/{\rm d}t} e^{i\epsilon {\rm d}t X_0 \sin [\omega(T-m{\rm d}t)]}\\
&\times e^{-i dt \sum_{j} \vec{\sigma}_{2j+1} \cdot \vec{\sigma}_{2j+2}}  e^{-i dt \sum_{j} \vec{\sigma}_{2j} \cdot \vec{\sigma}_{2j+1}} 
\end{aligned}
\end{equation}
where ${\rm d}t$ is some chosen time step, and $\vec{\sigma}_j = (X_j, Y_j, Z_j)$. The effect of the approximation on the DSF is better seen in the Lehmann representation \eqref{lehman}, where the eigenstates $|E_n\rangle$ are now those of the Floquet Hamiltonian $H_F$ defined as
\begin{equation}\label{floquet}
e^{-i{\rm d}t H_F} = e^{-i dt \sum_{j} \vec{\sigma}_{2j+1} \cdot \vec{\sigma}_{2j+2}}  e^{-i dt \sum_{j} \vec{\sigma}_{2j} \cdot \vec{\sigma}_{2j+1}}
\end{equation}
We show in Fig \ref{fig:time} the effect of the Trotter decomposition on the time evolution, by plotting the excitation energy and momentum of the eigenstates of the exact Hamiltonian and of the Floquet Hamiltonian. Especially on the lowest energy states, we observe only a relatively slight shift in energies.

The protocol to run on the quantum computer is thus the following. Firstly, we prepare the qubits in the ground state of the 1D Heisenberg model. We will explain in the next section how to implement this step. Secondly, we apply the operator $W(T)$ using the approximation \eqref{trotter}. The right-hand side of \eqref{trotter} can be implemented directly on a gate-based quantum computer, since it is a product of one-qubit gates and two-qubit gates. Thirdly, we measure all the qubits in the $X$ basis. From the measurement outcomes $X_j$, we compute $\hat{X}_q$ in \eqref{hatx}, obtaining an estimate for $\langle \hat{X}_q(T)\rangle_\epsilon$. Finally, we repeat the same experiment for $\epsilon\to -\epsilon$, and compute $S(q,\omega)$ from \eqref{dsf1d} at the finite value of $T$.

\begin{figure}
    \centering
    \includegraphics[width=0.9\linewidth]{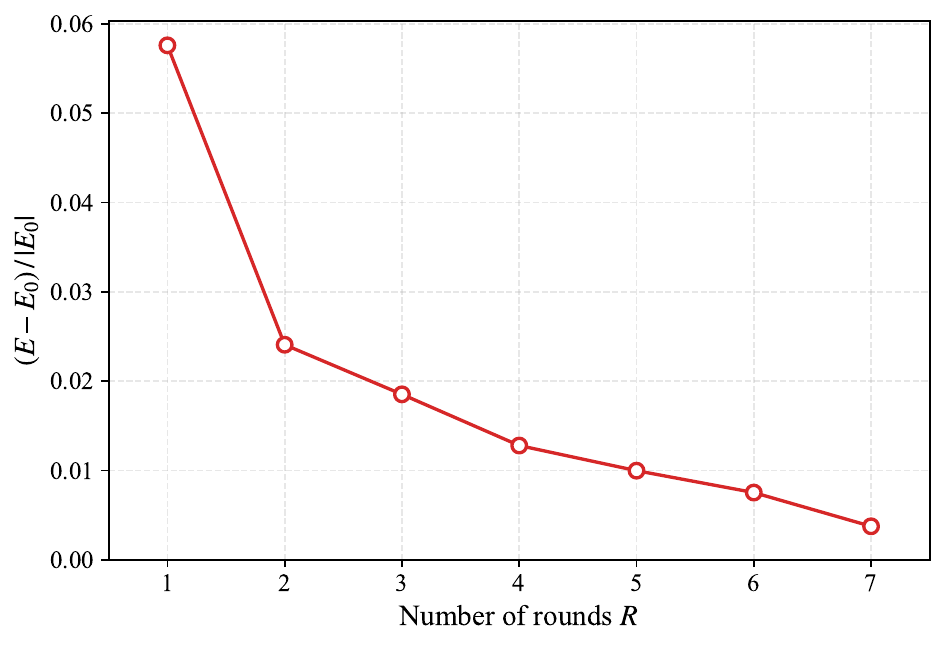}
    \caption{\emph{Ground state preparation.} Relative energy difference $(E-E_0)/|E_0|$ between the energy $E$ of the optimized ansatz~\eqref{ansatz} and the exact ground-state energy $E_0$ of the $L=20$ Heisenberg chain, as a function of the number of rounds $R$.
    }
    \label{fig:prep}
\end{figure}

\textbf{\emph{Preparation of the ground state of 1D Heisenberg model.}}--- Prior to time-evolving the qubits with the Hamiltonian \eqref{hamiltonian}, a good approximation of the thermal state must be prepared. We will focus on preparation of the ground state of the system at zero temperature, and set thus $\beta=\infty$ in \eqref{dsf1d}. To that end, we consider an adiabatic evolution \cite{kato1950adiabatic,albash2018adiabatic} interpolating between the staggered-field Hamiltonian
\begin{equation}
    H_{\rm stag}=\sum_{j=1}^L (-1)^j Z_j\,,
\end{equation}
and the Heisenberg Hamiltonian \eqref{hamiltonian}. The ground state of $H_{\rm stag}$ is the Neel state $|0101...\rangle$. Initializing the system in this Neel state, and time-evolving with a linear interpolation $(1-t/T)H_{\rm stag}+t/T H$ as a function of time $t$ for a total evolution time $T$, we observe that the energy of the final state at time $T$ converges quickly to that of the ground state of $H$. From this observation, we consider the following adiabatic-inspired ansatz
\begin{equation}\label{ansatz}
\begin{aligned}
   |\psi\rangle =\prod_{r=0}^{R-1} & e^{-i t_{4r+3}\sum_{j}(-1)^j Z_j} e^{-i t_{4r+2}\sum_j \vec{\sigma}_{2j+1} \cdot \vec{\sigma}_{2j+2}} \\
&e^{-i t_{4r+1}\sum_{j}(-1)^j Z_j} e^{-i t_{4r}\sum_{j} \vec{\sigma}_{2j} \cdot \vec{\sigma}_{2j+1}} |0101...\rangle
\end{aligned}
\end{equation}
for a given number of rounds $R$, analogously to the Quantum Approximate Optimization Algorithm \cite{farhi2014quantum}. At fixed number of rounds $R$, we optimize the parameters $t_r$ so as to minimize the energy of $H$. We present in Fig \ref{fig:prep} the minimal energy obtained as a function of the number of rounds $R$.

This exact parameter optimization can only be done in system sizes sufficiently small to be amenable to statevector simulations. However, these parameters obtained in size $L=20$ are also good parameters for the same ansatz in larger system sizes. Indeed, because of a lightcone effect, the energy density of the ansatz \eqref{ansatz} for $R$ rounds is constant with system size for $L\geq 2+4R$, because the lightcone of the circuit has not reached the boundaries of the system. So when optimizing in size $L=20$, the optimal parameters obtained for $R=1,2,3,4$ are also optimal in the thermodynamic limit $L\to\infty$.

\begin{figure}
    \centering
    \includegraphics[width=0.9\linewidth]{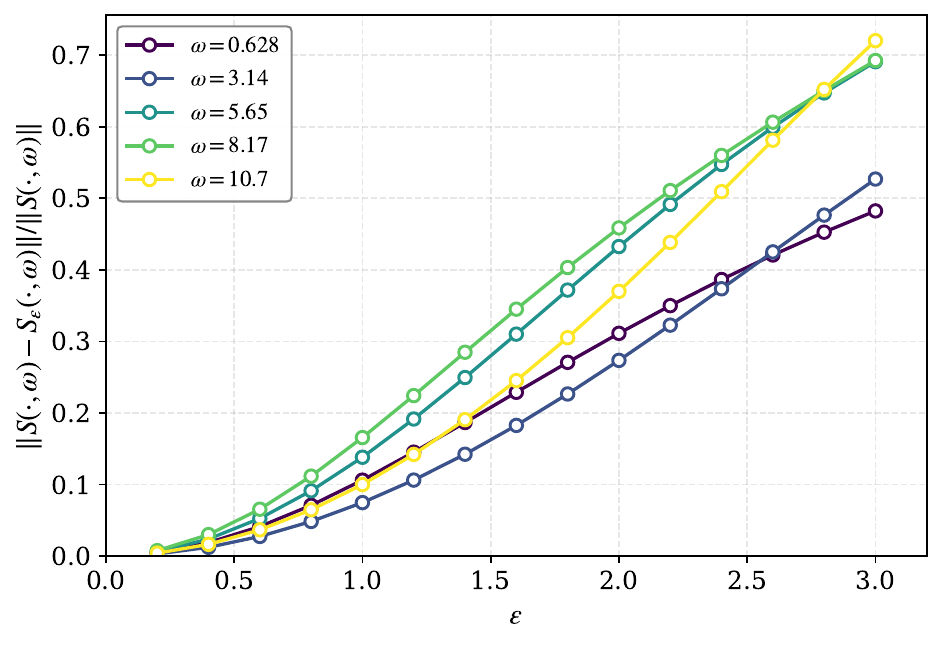}
    \caption{\emph{Influence of the coupling $\epsilon$.} Relative distance $||S(\cdot,\omega)-S_\epsilon(\cdot,\omega)||/||S(\cdot,\omega)||$ between the DSF in the limit $\epsilon\to0$ and the one computed at finite coupling $\epsilon$, as a function of $\epsilon$ for several target frequencies $\omega$. Here $S_\epsilon(\cdot,\omega)$ denotes the vector of momentum components $S(q,\omega)$ at fixed $\omega$, with $q=\frac{2\pi k}{L}$ and $k=0,\dots,L-1$, and $||\cdot||$ is the 2-norm. The data are obtained with the same parameters as the hardware implementation: $L=20$, $dt=0.24$, $T=2.4$.
    }
    \label{fig:epsilon}
\end{figure}

\textbf{\emph{Effect of the coupling $\epsilon$.}}--- Formula \eqref{dsf1d} for the dynamical structure factor is exact only in the limit $\epsilon\to 0$. On a quantum computer, this limit cannot be taken in practice. Indeed, the measurement of $\langle \hat{X}_q(t)\rangle_\epsilon$ inevitably comes with shot noise, of order $1/\sqrt{N_S}$ with $N_S$ the number of shots performed. Taking the difference appearing in \eqref{dsf1d} and dividing by $\epsilon$ amplifies this shot noise by a factor $1/\epsilon$. Hence, to take the limit $\epsilon\to 0$ at fixed precision, one has to scale the number of shots as $1/\epsilon^2$. On actual quantum computing hardware, this is prohibitive. The choice of $\epsilon$ results thus from a trade-off between lowering the shot noise by taking $\epsilon$ larger, and lowering the correction term $\mathcal{O}(\epsilon^2)$ in \eqref{dsf1d} by taking $\epsilon$ smaller. What value to take greatly depends on the model. In Fig \ref{fig:epsilon}, we plot the relative difference between the DSF in the limit $\epsilon \to 0$ and that computed with a finite coupling $\epsilon$. We see that setting $\epsilon=1$ gives around $10\%$ error, and $\epsilon=2$ around $30\%$ error.

\textbf{\emph{Hardware implementation.}}--- We now present a hardware implementation of the Fugaku--Reimei hybrid quantum--HPC system as the computational platform. The quantum circuit is executed on Quantinuum Reimei, while the classical optimization of the state preparation circuits is performed on Fugaku. The parameters of our experiment are set to $L=20$, Trotter step size ${\rm d}t=0.24$, Trotter steps $10$ to reach time $T=2.4$, and the coupling set to $\epsilon=2$. This value of coupling is chosen so as to increase the measured signal, while remaining in the region where it does not deform the DSF significantly. The ground state is prepared by using only one round $R=1$ of the ansatz \eqref{ansatz}. We measure the following values of $\omega$
\begin{equation}
    \omega=\frac{2\pi n {\rm d}t}{T}\,,\quad n=1,2,3,...,20\,.
\end{equation}
We perform $2000$ shots for each setting. Every circuit contains $220$ so-called TK2 gates, which are sequential $XX$, $YY$, and $ZZ$ rotations with arbitrary gate angles.

To quantify the deviation in $S(q, \omega)$ due to the adiabatic approximation, we calculate
$
S_\mathrm{diff} = \sum_{q, \omega} |S(q, \omega) - S_0(q, \omega)|^2 / \sum_{q', \omega'} |S_0(q', \omega')|^2,
$
where $S_0(q, \omega)$ is obtained using the exact ground state, fixing $dt=0.24$, $T=2.4$, and $\epsilon=2$.
Surprisingly, using only the first layer of the $R=2$ optimal adiabatic evolution yields a smaller $S_\mathrm{diff} (\simeq 0.0381)$ than the full $R=2$ evolution $(\simeq 0.0568)$.
Although this is a special case of the parameter set, we used only the first layer to further reduce the number of gates.

\begin{figure}
    \centering
    \includegraphics[width=0.85\linewidth]{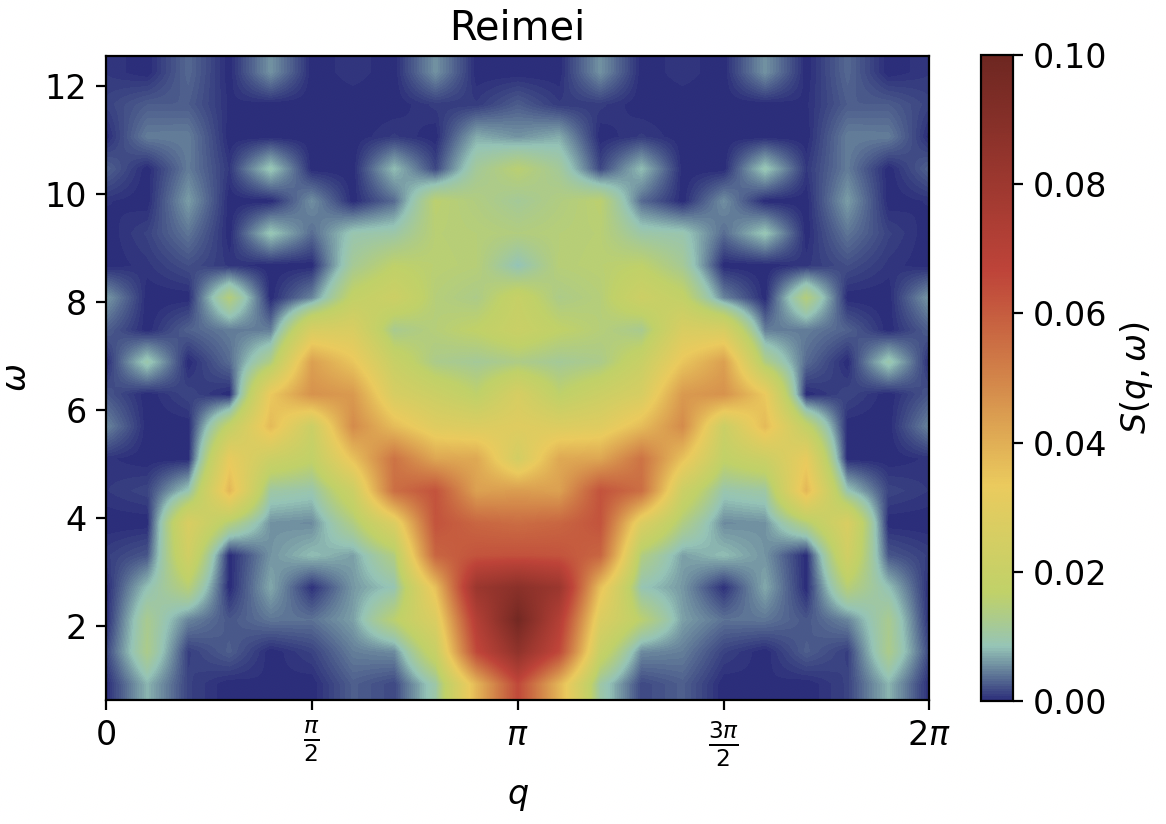}
    \includegraphics[width=0.85\linewidth]{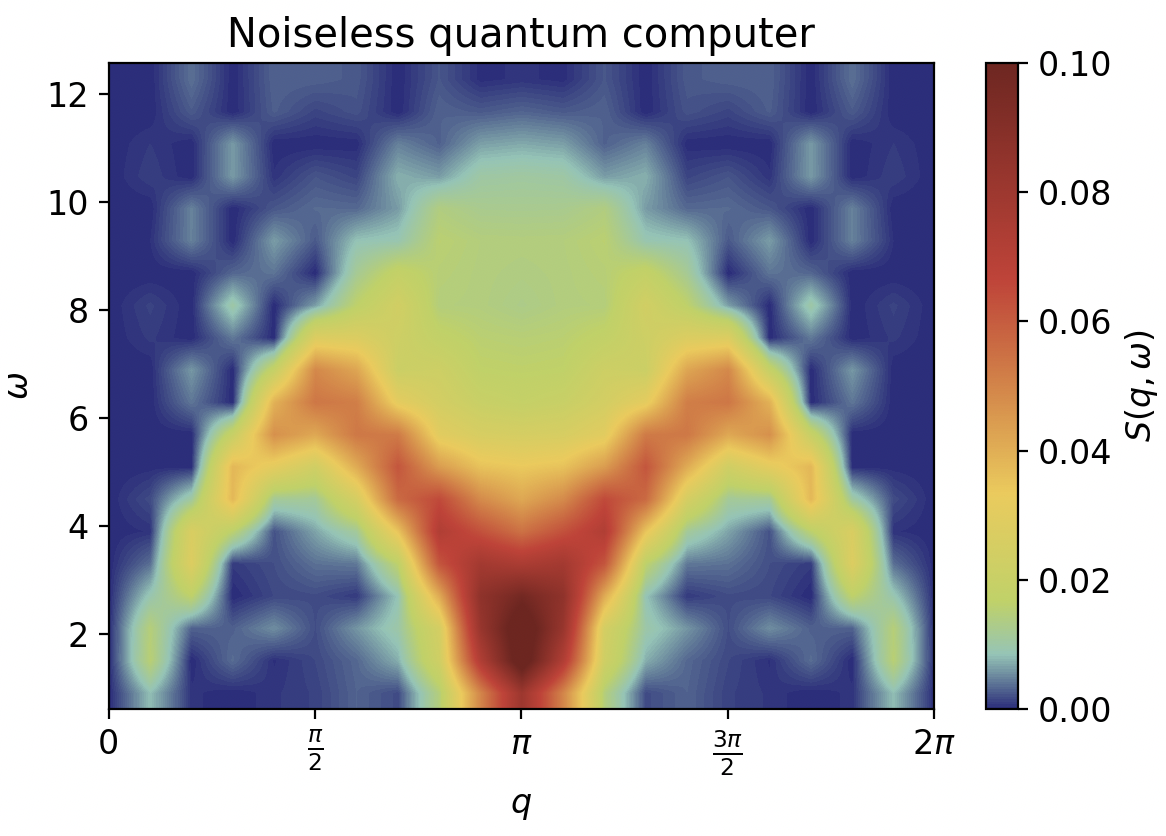}
    \includegraphics[width=0.87\linewidth]{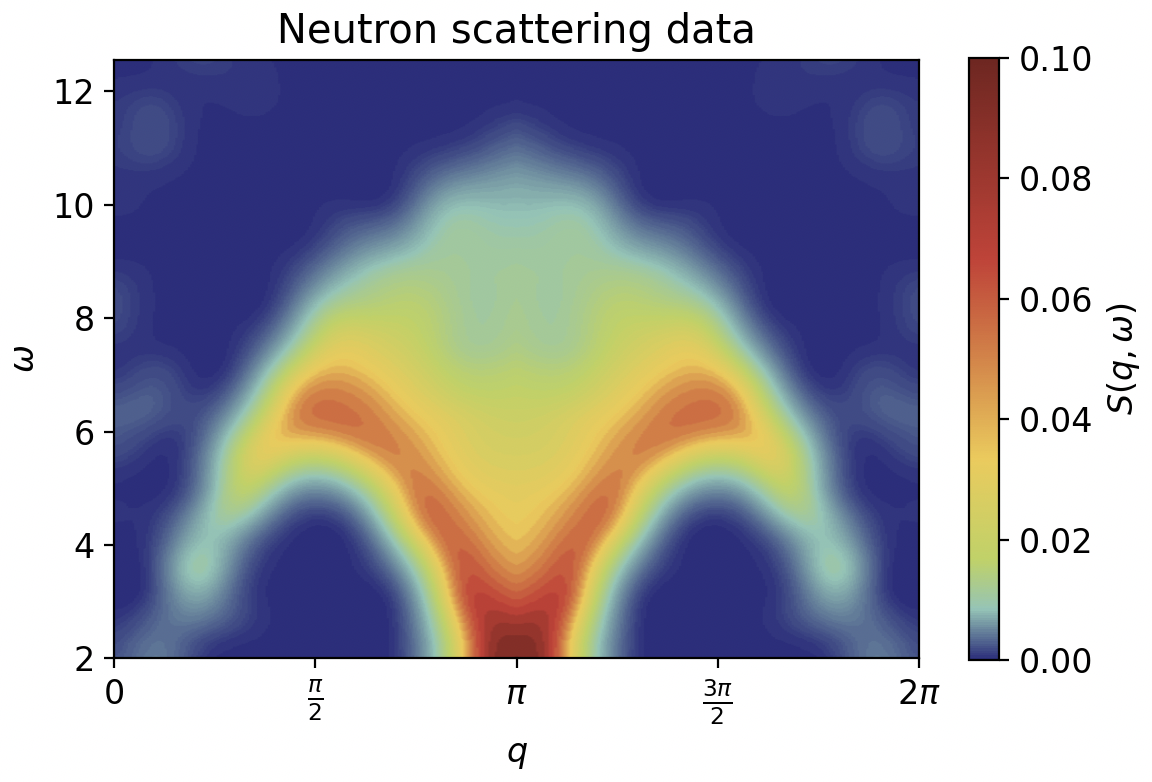}
    \caption{\emph{Comparison between DSFs.} Color plot for $S(q,\omega)$, as measured on Quantinuum Reimei ion-trapped quantum computer (top panel), on an ideal noiseless quantum computer using the same circuits (middle panel), and obtained from neutron scattering experiments (bottom panel). The data for the neutron scattering experiment panel was numerically extracted from the plots of Ref. \cite{mourigal2013fractional}, with the overall arbitrary scale of $S(q,\omega)$ being chosen so that the maximal value is $0.1$. }
    \label{fig:reimei}
\end{figure}

We display a 2D color plot of the DSF $S(q,\omega)$ measured on Quantinuum Reimei in Fig \ref{fig:reimei}. We compare the results with the noiseless values obtained from the same circuits without shot noise, and with neutron scattering experiments of the copper sulfate crystal, whose data is numerically extracted from the plots of Ref. \cite{mourigal2013fractional}. We see excellent agreement between the quantum computing hardware experiment and noiseless simulations of the same circuits. This agreement is obtained without any noise mitigation. The DSF obtained agrees also very well with neutron scattering experiments.

\textbf{\emph{Discussion.}}---We have introduced an out-of-equilibrium driving Hamiltonian approach to compute the DSF of spin models on quantum computers. We have tested the method on the Quantinuum Reimei ion-trapped quantum computer and obtained very good agreement with noiseless values, as well as with the copper sulfate crystal. Unlike previous methods~\cite{lee2026benchmarkingquantumsimulationneutronscattering}, we directly simulate a toy model version of a neutron scattering experiment, rather than computing dynamical correlations in real time. In practice, this allows us to directly target specific values of $\omega$. 

Beyond neutron scattering experiments, this protocol and the ability to target specific frequencies can be relevant for example for Nuclear Magnetic Resonance simulations \cite{marthaler2025good,khedri2024impact,seetharam2023digital,burov2024towards,burov2025large}. Indeed, the spectra of molecules obtained there from the free-induction decay (that is mathematically similar to the DSF studied in this manuscript) are typically sharply localized at specific values of frequencies, while involving long evolution times. Such protocol would bring significant savings in runtime, by using just a few circuits at fixed values of frequencies, instead of having to sample the free-induction decay over a long stretch of time.

While our pumping approach to compute the DSF of a given state is scalable, the state preparation that we implemented in this work involved classically optimised ansatz circuits using exact statevector methods. By definition, no quantum advantage is possible in this regime. However, there are several options to scale the state preparation to larger system sizes using the HPC resources. For example, classical methods based on Sparse Pauli Dynamics have been used to obtain approximate ground states for system sizes beyond the exact state vector regime~\cite{granet2025superconducting, Lin2026utilityscalequantum}. 
Alternatively, if the ground state can be obtained classically as a matrix product state, for instance using the density-matrix renormalization group~\cite{white1992density,schollwock2011density}, it can be embedded into a shallow quantum circuit~\cite{schon2005sequential,ran2020encoding,rudolph2024decomposition,malz2024preparation}. This is expected to be particularly effective and scalable for one-dimensional systems. Combining these recent methods with our DSF algorithm within the hybrid quantum-HPC framework may lead into a regime where the joint task of preparing the initial state and computing the structure factor becomes challenging for classical computers alone.

\textbf{\emph{Acknowledgments.}}---We thank Alec Owens and Eli Chertkov for feedback on the manuscript. Part of this work is based on results obtained from a project, JPNP20017, commissioned by the New Energy and Industrial Technology Development Organization (NEDO).
This work used computational resources of Fugaku provided by RIKEN Center for Computational Science (Project ID: ra010014).
H.S. acknowledges support from the Center of Innovation for Sustainable Quantum AI (SQAI), JST under Grant No. JPMJPF2221.


%

\end{document}